\documentclass{article}
\usepackage{amssymb,amsmath,bm}
\begin{document}
\begin{titlepage}
\noindent{\large
\textbf{On non-parallelism of spin and magnetic moment of proton}}
\vspace{\baselineskip}
\begin{center}
Amir~H.~Fatollahi~{\footnote {ahfatol@gmail.com}}\\
\vspace{\baselineskip}
\textit{Department of Physics, Alzahra University, Vanak 19938-93973, Tehran, Iran}
\end{center}
\vspace{\baselineskip}
\begin{abstract}
\noindent The nonrelativistic quark model predicts that the spin and the magnetic 
moment of proton are not parallel. Based on, the modifications are outlined that 
sounds crucial for the analysis of the data out of scattering off the protons.
\end{abstract}

\vspace{1cm}

\end{titlepage}

\noindent According to the quark model the magnetic moment of proton is
the vector sum of the magnetic moments of its three constituent
quarks \cite{griff,halzen}
\begin{equation}\label{1}
\bm{\mu}=  \bm{\mu}_1+ \bm{\mu}_2+ \bm{\mu}_3
\end{equation}
in which
\begin{equation}\label{2}
\bm{\mu}_a=\frac{Q_a}{m_a c}\mathbf{S}_a,~~~~~~~~~~~a=1,2,3
\end{equation}
with $Q_a$, $m_a$ and $\mathbf{S}_a$ representing the electric charge, the constituent
mass and the spin operator of the $a\;\!$th quark, respectively. On the other hand,
the state of a proton is usually taken such that is a simultaneous eigen-state of the
square and the $z$-component of the operator
\begin{equation}\label{3}
\mathbf{S}=\mathbf{S}_1+\mathbf{S}_2+\mathbf{S}_3
\end{equation}
with eigen-values $\frac{3}{4}\hbar^2$
and $\pm\frac{1}{2}\hbar$, respectively.
For example, the spin-up proton state is presented by \cite{griff}
\begin{align}\label{4}
|\mathrm{p}\!\uparrow\rangle =\frac{1}{3\sqrt{2}}\Big(
&|\mathrm{udu}-\mathrm{duu}\rangle
|\!\uparrow\downarrow\uparrow - \downarrow\uparrow\uparrow\rangle \cr
+
&|\mathrm{uud}-\mathrm{udu}\rangle
|\!\uparrow\uparrow\downarrow - \uparrow\downarrow\uparrow\rangle\cr
+
&|\mathrm{uud}-\mathrm{duu}\rangle
|\!\uparrow\uparrow\downarrow - \downarrow\uparrow\uparrow\rangle\Big)
\end{align}
in which $|\mathrm{u}\rangle$ and $|\mathrm{d}\rangle$ represent the
up and down flavors of quarks, respectively. A similar expression can
be given for the spin-down proton state, $|\mathrm{p}\!\downarrow\rangle$.
The point is, one can easily check that neither $|\mathrm{p}\!\uparrow\rangle$ nor
$|\mathrm{p}\!\downarrow\rangle$ are eigen-states of the
$z$-component of the vector (\ref{1}). The reason simply backs to the
fact that, even by equating the constituent masses $m_a$'s, the
electric charges of up and down flavors are different. By the same reason,
and as it is evident by (\ref{1}) and (\ref{3}),
the spin and the magnetic moment of proton are not parallel.
The commutation relations
\begin{align}
\label{5}
[\mu_i,S_i]&=0,  \\
\label{6}
[\mu_i, \mathbf{S}\cdot\mathbf{S}]&\neq 0,\quad i=x,y,z
\end{align}
mean that, for example, the $\mu_z$ and $S_z$ can be determined
simultaneously, but not in the basis in which the total-spin is given.

This simple observation above should be contrasted with the widely used
expression that the relation between the spin and magnetic moment
of proton is given by \cite{griff,halzen,leader}
\begin{equation}\label{7}
\bm{\mu}_\mathrm{p} =g_\mathrm{p}\frac{e}{2\,M_\mathrm{p}}\,\mathbf{S},
\qquad g_\mathrm{p}\simeq 5.59
\end{equation}
saying that, 1) the spin and magnetic moment of the proton are parallel, 2) the
total spin squared ($\mathbf{S}\cdot\mathbf{S}$), as well as the $z$-components
of the total-spin and the total magnetic moment ($S_z$ and $\mu_z$, respectively)
can be determined simultaneously.

The clarification of the above mentioned mismatch between the quark model and
the relation (\ref{7}), together with the domain of validity of (\ref{7}),
are presented in \cite{fsm06}. In particular, it is shown
that the relation (\ref{7}) should be interpreted as a relation between
the matrix elements of the spin and the magnetic moment in the lowest order of
perturbation, where transition between different baryon states does not occur.
In fact a direct calculation shows that
\begin{align}\label{8}
\langle \mathrm{p},s_z| \bm{\mu}\cdot\mathbf{B} | \Delta^+, s_z\rangle \neq 0,
\qquad& \mathbf{B}\parallel\mathbf{\hat{z}} \\ \label{9}
\langle \mathrm{p},s_z| \bm{\mu}\cdot\mathbf{B} | \Delta^+, s'_z\rangle \neq 0,
\qquad& \mathbf{B}\perp\mathbf{\hat{z}}\;\&\;s_z\neq s'_z.
\end{align}
in which $\Delta^+$ is the spin-3/2 ``uud" state, and $\bm{\mu}$
is given by the quark model, (\ref{1}).
Hence if the energy due to the coupling of the magnetic moment with
the external field is so high that transition from one baryon state to
another one is possible, the relation (\ref{7}) is not valid anymore.
The threshold of the magnetic field
by which the above matrix elements find significant contribution is
$B\sim 3\times 10^{19}$~Gauss. Though this threshold is extremely high, interestingly
in the so-called deep inelastic scattering of electrons off the
protons this threshold is quite accessible \cite{fsm06}.

It seems crucial to consider the consequences of above remarks
on the analysis of data out of the deep inelastic scattering
off the protons. To have a better insight, let us begin with
the case with elastic scattering experiments.

Let us start with a Dirac point particle such as electron. The electron couples to
the electromagnetic potential through the term $-e\, j\cdot A$, in which the
current in momentum space is given by \cite{halzen} (units $\hbar=c=1$)
\begin{align}\label{10}
j^\mu & = \bar{u}(k')\, \gamma^\mu\, u(k)\cr
& =\! \frac{1}{2m_\mathrm{e}} \bar{u}(k')\Big( (k+k')^\mu + i \sigma^{\mu\nu} (k'-k)_\nu\Big) u(k)
\end{align}
with the second line known as the Gordon decomposition of the current.
The second term of the Gordon representation is responsible for the spin dependent
transitions. For example, in the nonrelativistic limit and for a time independent
potential, it can be shown that the coupling of the second term reduces to \cite{halzen}
\begin{equation}\label{11}
\psi^\dagger \Big( \frac{e}{2m_\mathrm{e}} \bm{\sigma}\cdot \mathbf{B} \Big) \psi
\end{equation}
which is nothing but the magnetic dipole coupling of electron to
the magnetic field, $-\bm{\mu}_\mathrm{e}\cdot\mathbf{B}$,
with $\bm{\mu}_\mathrm{e} = -\frac{e}{m_\mathrm{e}} \mathbf{S}_\mathrm{e}$.
In above $\psi$ is the two-component wave-function of electron.

In the elastic scattering off protons, inspired by (\ref{10}),
the current for the proton is taken \cite{halzen}
\begin{equation}\label{12}
J^\mu = \bar{u}(p')\Big( F_1(q^2) \,\gamma^\mu
+ \frac{\kappa}{2M_\mathrm{p}}\,F_2(q^2)\, i \sigma^{\mu\nu} q_\nu \Big) u(k)
\end{equation}
in which $q\equiv p'-p$ is the change of the 4-vector momentum of proton.
The functions $F_1$ and $F_2$ are known as the form factors of the proton, and $\kappa$
is sitting in account of the anomalous magnetic moment of the proton. In the low momentum
transfer limit, $q^2\to 0$, one expects $F_1(0)=F_2(0)=1$, that is the proton acts
like a structure-less particle. Then in the limits mentioned for the electron,
the coupling of the proton to the magnetic field is reduced to (units $\hbar=c=1$)
\begin{equation}\label{13}
(1+\kappa) \, \frac{e}{2M_\mathrm{p}} \bm{\sigma}\cdot \mathbf{B}
\end{equation}
which, by comparison to (\ref{5}), gives $\kappa=-1+ g_\mathrm{p}/2 \simeq 1.79$.
Now the point is, as mentioned, due to non-parallelism of magnetic moment and spin of the proton,
the coupling of the above form for proton is only valid in an averaged sense, which
is not expected generally be sufficient in an experiment set-up by which the sub-structure
of proton is probed. So one may try to alter the initial form (\ref{12}), as well as
replacing the incoming $u(p)$ and outgoing $\bar{u}(p')$ wave-functions by a composite form like
(\ref{4}), such that in the mentioned limit the resulting magnetic moment come in the form suggested by
the quark model, (\ref{1}).

Based on above, the situation with inelastic experiments needs a more drastic change
in point of view. According to the quark model picture of hadrons, the magnetic
moment and spin of proton are due to its constituent quarks, relations (\ref{1}) and (\ref{3}).
Contrary to the expectation about spin, the present analysis given for the data out of the
polarized deep inelastic scattering experiments has allowed just about a third to the contribution of
valence quarks to proton's spin \cite{bass}.
The basis for the modifications outlined here is grounded on the fact that, if the result of an
analysis is going to be compared with the quark model expectations, it is logical
that the starting point would be chosen at most compatibility place with the quark model too.

The differential cross-section to find the scattered
electron in the solid angle $d\Omega$ and energy range $(E', E'+dE')$ is given by \cite{leader}
\begin{equation}\label{14}
\frac{d^2\sigma}{d\Omega dE'} = \frac{\alpha^2}{2M_\mathrm{p} \,q^4}
\frac{E'}{E} \,L_{\mu\nu} \,W^{\mu\nu}
\end{equation}
The spin dependent sector of the leptonic tensor $L_{\mu\nu}$, which is its antisymmetric
part in $\mu$ and $\nu$, is
\begin{equation}
L^{(\mathrm{A})}_{\mu\nu} = 2\,i m_\mathrm{e}\, \epsilon_{\mu\nu\alpha\beta} \, s^\alpha \, q^\beta
\end{equation}
which contracts with the antisymmetric part of the hadronic tensor $W_{\mu\nu}$
\begin{align}
\frac{1}{2M_\mathrm{p}} W^{(\mathrm{A})}_{\mu\nu} = &~
\epsilon_{\mu\nu\alpha\beta} \, q^\alpha \Big\{ M_\mathrm{p}\, \mathcal{S}^\beta G_1(\nu,q^2)
\cr
& +  \left[(P\cdot q) \, \mathcal{S}^\beta - (\mathcal{S}\cdot q) P^\beta\right]
\frac{G_2(\nu,q^2)}{M_\mathrm{p}} \Big\}
\end{align}
with $s^\alpha$ and $\mathcal{S}^\beta$ as the covariant spin vector of
electron and proton, respectively, and $P^\alpha=(M_\mathrm{p},\mathbf{0})$.
As is evident, the expression chosen for $W^{(\mathrm{A})}_{\mu\nu}$ is
such that in the appropriate limit can produce solely the coupling of
the spin of proton to external probe. In other words,
by this choice there is no term in account
of the direct coupling of the electron with an individual ``bared" or
``screened" quark; a phenomena that supposedly
one expects in the set-up for deep inelastic experiments.
As mentioned, one may try to suggest an altered form for
$W^{(\mathrm{A})}_{\mu\nu}$ which would be more compatible with quark model.
By assigning to each individual constituent quark its own form factor, in a static
quark model picture of proton, one may suggest
\begin{align}
\frac{1}{2M_\mathrm{p}} W^{(\mathrm{A})}_{\mu\nu}= &~ \epsilon_{\mu\nu\alpha\beta}
\sum_{a=1}^3  z_a q^\alpha\Big\{ M_\mathrm{p}\, S_a^\beta G_{a1}(z_a;\nu,q^2)
\cr
 &~+  z_a \left[(P\cdot q) \, S_a^\beta - (S_a\cdot q) P^\beta\right]
\frac{G_{a2}(z_a;\nu,q^2)}{M_\mathrm{p}} \Big\}
\end{align}
in which $S_a^\alpha$ is for the covariant spin vector of the $a\;\!$th quark,
and $z_a$ is the fraction of the transferred momentum $q^\mu$ which has been shared
to $a\;\!$th quark in a single process, hence $0\leq z_a \leq 1$ and $\sum z_a=1$.
The dependence of form factors are assumed as
\begin{equation}
G_{ai}(z_a;\nu,q^2) = G_{ai}(z_a\nu, z_a^2q^2),~~~~~i=1,2
\end{equation}
If one of $z_a$'s approaches one, the scattering occur as if it is a scattering
off one of the constituent quarks. So
\begin{equation}
G_{ai}(z_a;\nu,q^2) \to 0 ~~~~\mathrm{for}~~~ z_a\to 0
\end{equation}
Also, by letting that each constituent quark has its own form factor, one in
fact lets that the coupling with the magnetic moment suggested by quark model,
relation (\ref{1}), would occur.
Further, it is assumed that the form factor of each constituent quark
obeys the Bjorken scaling
\begin{equation}
G_{ai}(z_a;\nu,q^2) \to G_{ai}\left(\frac{\nu}{q^2}\right), ~~~\mathrm{for~large}~``-\!z_aq^2"
\end{equation}
By these all together, one expects that even a single constituent quark can regenerate
the data in the extreme deep inelastic regime, Bjorken's $x\to 0$, by which one can just
explain about a third of the proton spin.
\\[\baselineskip]
\textbf{Acknowledgement}:
The author is thankful for helpful discussions with M.~Khorrami and A.~Shariati on the
issue of the magnetic moment in the context of quark model. This work is partially
supported by the Research Council of the Alzahra University.

\end{document}